# A Survey on the Adoption of Cloud Computing in Education Sector

Rania Mohammedameen Almajalid

College of Computing and Informatics, Saudi Electronic University, Jeddah, Saudi Arabia

Seidenberg School of Computer Science and Information Systems, Pace University, Pleasantville, New York

`r.almajalid@seu.edu.sa`

**Abstract**—Education is a key factor in ensuring economic growth, especially for countries with growing economies. Today, students have become more technologically savvy as teaching and learning uses more advance technology day in, day out. Due to virtualize resources through the Internet, as well as dynamic scalability, cloud computing has continued to be adopted by more organizations. Despite the looming financial crisis, there has been increasing pressure for educational institutions to deliver better services using minimal resources. Leaning institutions, both public and private can utilize the potential advantage of cloud computing to ensure high quality service regardless of the minimal resources available. Cloud computing is taking a center stage in academia because of its various benefits. Various learning institutions use different cloud-based applications provided by the service providers to ensure that their students and other users can perform both academic as well as business-related tasks. Thus, this research will seek to establish the benefits associated with the use of cloud computing in learning institutions. The solutions provided by the cloud technology ensure that the research and development, as well as the teaching is more sustainable and efficient, thus positively influencing the quality of learning and teaching within educational institutions. This has led to various learning institutions adopting cloud technology as a solution to various technological challenges they face on a daily routine.

## 1 INTRODUCTION

THE pressure to deliver more quality services from very minimal resources has engulfed learning institutions across the globe as a result of the ongoing financial crises. But if resources, including IT services are shared among learning institutions, then there is a better chance that such institutions can have a better concentration on research, as well as core academic activities.

Additionally, the education system has witnessed gradual expansion thus transforming a learner centered system rather than teacher centered system. The new leaning environment calls for innovate learning, as well as teaching tools. As more tools are provided by new technology with each day that goes by, the education system continues to evolve more rapidly. E-learning and virtual classrooms have become the order of the day in most learning institutions whereby all the consultants can access the education system, monitor and even provide guidance accordingly.

Today, the word cloud computing is an essential terminology in the IT world and has been applied in different sectors including the educational sector. Cloud computing can be defined as the use of virtual resources that are highly scalable and can be shared by different and diverse users. Cloud computing was driven by the increased use of electronic devices including laptops, smart phones and tablet PCs. Current studies have indicated that this is among the fastest growing areas within the digital economy.

Cloud computing can be termed as providing information technology resources when demanded through the Internet. Cloud computing is of various forms including public cloud, private cloud, as well as the option of "cloud as a service" as explained later in this research paper. The fact that cloud computing has evolved over time has provided enormous and versatile amount of resources that are scalable in nature depending on the needs of business in addition to serving large interests for potential, as well as immediate returns on the investments. By incorporating cloud computing in the IT strategy, institutions are able to increase their overall capacity while at the same time maintaining the required security level and limited infrastructural investment, hence ensuring that the total cost of ownership (TCO) is low. Thus, it is of essence that educational institutions to get better ways to provide inexpensive services of good quality using academic tools and research practices by striking a balance between cloud and on-premise services. Through a combination of best practices in grid computing, web technologies, virtualization, as well as utility computing, the cloud computing results as a computing infrastructure which carries with it the virtualization agility and the scalability associated with grid computing, as well as the simplicity that comes with Web 2.0. Despite the many benefits associated with cloud computing, it also has various disadvantages that emanate from the fact that all applications and data are situated within the internet. Not only does cloud computing provide teachers and students with free access to different services and applications within the cloud for both informal and formal use in the education sector, but it also provides greater mobility and flexibility when it comes to the utilization of resources and creates a learning environment that is personalized or rather virtual teaching and learning. Cloud computing ensures that trainers, staff, as well as students can access any form of information from anywhere, using any type of device. Thus, both private as well as public learning institutions can adopt this type of technology as a way of providing improved services using the few resources at their disposal.

The use of cloud computing in institutions of higher learning has provided many benefits to universities and colleges. First, cloud computing is cost effective thus efficient where the resources are minimal. Secondly, the academicians derive a positive experience from its use. Additionally, the IT staffs productivity is increased by the use of cloud computing. In todays world, various cloud platforms including Google and Microsoft are providing various application and services free of charge to staff and students in different learning institutions. Some examples of these services which provided free of charge; calendars, emails, document storage, contact list, website creation services, as well as document sharing services.

The remainder of this survey is organized as follows: In the next section, the basic background of cloud computing has been given including the three service models, deployment models, as well as the features, benefits and concerns of cloud computing. Usage of cloud computing in education is discussed in Section 3, cloud computing in e-learning is presented in Section 4. Impact of cloud computing on the Saudi educational sector and future work to move into cloud computing for Saudi Arabia institutions are provided in Section 5 and Section 6, respectively. Finally, the conclusion is given in Section 7.

## 2   CLOUD COMPUTING

There several definitions of the term cloud computing are existing. According to Google, this term refers to open standards that are service based, safe, convenient, internet centric, and fast network computing and data storage services. But IBM describes cloud computing as pooling computing resources that are virtual in nature and form the new era of IT. Despite the difference in the definitions, it is evident that the term refers to a super computing model that is internet-based.

Cloud Computing can be simply defined as a pool of computing resources that are delivered through the web. If a diagram representing the relationships among the various elements, it takes the shape of a cloud. It takes into account all applications, networks, as well as servers. Thus, the end user is able to access all these elements using the internet. Cloud computing allows the consumer to pay for only few things including what they use like the processing time, memory and the bandwidth, hence making it cheaper.

Cloud computing provides value that is dissimilar to the traditional It environment. It offers economies of scale through aggregating computing resources and virtualization. Cloud computing ensures a global reach of information and services using a computing environment which allows on-demand scalability and minimal initial investment. It can also provide pre-built services and solutions together with the required skills for running and maintaining them, hence reducing the risk and need for the institution to maintain a group of staff that is highly skilled and scarce to find.

As for the end-users, they do not have to purchase new equipment or maintain hardware, upgrade or update existing software, obtain licenses for the software or data synchronization since the clouds service includes all of them. Cloud computing is highly scalable thus the quantity of resources used depends on the application needs and payment is based on the actual resources used. Further, it is characterized by mobility, and well as independency of the platform meaning that it can be accessed from anywhere, anytime and from any device.

### 2.1   Cloud Service Models

It is important to understand the different service types offered by cloud computing in a bid to understand cloud computing as a new approach to IT. Discussed below are the various service types offered through cloud computing.

1) **Infrastructure as a Service  (IaaS)**
   IaaS is responsible for various aspects including running the application and operating systems, housing, maintaining and operating the various equipment on behalf of the client. However, IaaS cannot manage the underling cloud infrastructure. Payment by consumers is based on utility computing basis. Some of the characteristics associated with IaaS include dynamic scaling, internet connectivity, administrative tasks that are automated, platform virtualization, as well as lower total ownership costs leading to lower capital. Some of the IaaS offered by dealers include Rackspace Cloud Servers, Google, Amazon EC2, as well as IBM.

2) **Software as a Service (SaaS)**

SaaS ensures that clients are able to utilize the various providers applications that run on the cloud infrastructure, but are not in a position to control its hardware, network infrastructure or the operating system. It ensures access to applications that are functional in nature including CRM, web conferencing, the ERP, as well as the email among other applications. SaaS is associated with various benefits including security, rapid scalability, software compatibility, global accessibility, as well as reliability. It is also responsible for different business operation tasks including human resource management, content management, accounting, as well as computerized billing among others. Some of the providers that offer SaaS include NetSuite, Google, Citrix and Salesforce.com among others.

3) **Platform as a Service (PaaS)**

PaaS enables the customer to hire virtual servers, as well as other services required to operate the applications that exist. Further, it ensures that the client design, develop, test, deploy and host applications. Clients can deploy and control applications e.g. the configurations of the hosting environment, but they are not in a position to control the hardware, operating system, as well as the network infrastructure. Some of its characteristics include lack of software upgrades, reduced risk, and simplified deployment. The providers of PaaS include Google App Engine Salesforce.com, Microsoft Azure, as well as Rackspace Cloud Sites.

## 2.2 Cloud Deployment Models

The hosting models for cloud represent the different types of cloud environments which are characterized by different sized, access and proprietorship. There are four deployment models which are discussed below;

**Public Cloud:** In this type of cloud hosting, the cloud services are provided through a network that is accessible by the public. This model is perceived to be the ideal illustration of cloud hosting. In this type of hosting the provider offers infrastructure and services to a wide range of clients.

**Private Cloud:** This is also regarded as an internal cloud. The environment on which the cloud computing platform lies is protected by a firewall that is monitored by the information technology department which belongs to the particular organization and can only be used by the authorized clients only.

**Community Cloud:** This denotes a cloud hosting that is mutual and is shared among many organizations of a specific community including trading firms, banks, or gas stations among others. The group of users must have computing apprehensions that are similar.

**Hybrid Cloud:** This is an integrated model of cloud computing environment. It may consist of two cloud servers or more, which may either be public, communal or private. The servers are interconnected though each remains as a separate entity. A hybrid cloud is advantageous because it can overcome boundaries and cross isolation by the provider, but cannot be categorized among the public, communal or private clouds.

## 2.3 Cloud Computing Features

Cloud computing has been so successful because of its simplicity when it comes to its usage, as well as the several other advantages that accrue from its use. It is a cost effective solution for enterprises when it comes computing operations. Cloud computing is characterized by various features including;

1) Optimal Server utilization: Cloud computing ensures that servers are optimally utilized.
2) On-demand cloud services ensure that the customer is furnished with a tailor-made environment that is customized according to the needs of the client.
3) Dynamic Scalability: Cloud computing is a source of an additional processing buffer with no extra capital investment by the users.
4) Disaster Recovery: This is one of the core feature associated with cloud computing. It mitigates the need of comprehensive disaster recovery plans for the information technology infrastructure. It ensures faster recovery and information is available in multiple sites hence making it more efficient and effective.
5) Virtualization technology: Virtualization is considered to be the most essential cloud computing feature. This means that the computing elements are not real; rather, they are of a virtual nature. It is possible for the virtualization technology to expand the hardware capacity thus simplifying the reconfiguration process of the software. The virtualization technology ensures that the platform can run several operating systems, with each application running independently as a result of the single-CPU simulation of multi parallel CPUs.

## 2.4 Benefits of Cloud Computing

The popularity of could computing is growing day by day. Many organizations are adopting cloud computing due to its numerous benefits. If cloud computing is adopted in the firm, the need to employ highly qualified personnel in the IT department is reduced. Discussed below are some of the advantages that accrue from cloud computing:

1) Convenience and improved accessibility: It can be accessed from anywhere and with any device by end-users. Further, it is convenient and easy to use. The user is not required to download any software or upgrade data to cloud in a bid to use. The only requirement is a device that can access internet and the internet itself.

2) Cost saving: The costs of installing IT infrastructure are always high and are considered to be capital expense. Through resource pooling, these costs are averted in cloud computing. Optimal use of software and hardware is achieved through resource pooling, hence increasing the efficiency and effectiveness of the available resources.

3) Reduced expenditure on Technology Infrastructure: The upfront spending on IT infrastructure is minimized while at the same time easy access to information is still ensured. Payment is on demand basis.

4) Minimal training on the personnel: Fewer people are required to perform more tasks on cloud. The level of skill regarding software and hardware issues is also minimal, hence easy to deploy and develop.

5) Super-Computing Power: Several computers are brought together to constitute a super server in cloud computing, hence providing users with a powerful data processing and computing capacity.

## 2.5 Cloud Computing Concerns

Despite the advantages associated with cloud computing, there are also various drawbacks which should also be considered. The most critical concerns relate to vendor lock, latency, reliability, security, control, performance, as well as privacy. Organizations may hesitate from surrendering the control of their IT resources to external providers may end up changing the existent technology without the consent of the customers. Thus, because the users do not have any control over the servers, they depend on the provider to manage and update their software.

Additionally, there exist valid privacy and security concerns. Cloud service implementation on a large scale may not be possible until legal matters relating to data protection and privacy are addressed. Vendor lock, as well as failure is also another concern of cloud computing. Proprietary APIs are used by several cloud providers to proffer their services. As the number of providers increase, portability is bound to become more important. Further, if a provider who owns the data center where a user has saved his/her data fails, there ought to be adverse repercussions to the user.

Finally, reliability is also another problem facing people and firms using cloud. For example, in February 2008, Salesforce.com was out of service for at least 6 hours, while Amazons EC2 (Elastic Compute Cloud) and Amazons S3 (simple storage service) experience an outage for 3 hours in the same month, and later in July 2008, S3 it experienced an outage for 8 hours [3]. In 2009, Gmail, a webmail service by Google was out for 3 hours leaving more than 110 million users stranded [4].

## 3 USAGE OF CLOUD COMPUTING IN EDUCATION

Cloud computing ensures that learning institutions do concentrate more on research and learning, rather than on implementing complex IT infrastructure. The cloud computing applications related to education will form the basis of future IT infrastructure in education to ensure the development of hardware and software environment. By integrating the resources through cloud computing, it will be possible to meet the high demand by utilizing the high speeds involved in processing the data thus reducing the pressure associated with the information explosion. This will further ensure development in education by utilizing the speedy changes in IT, as well as the software as a service SaaS (trend). It will do so by enhancing the kind of resources utilized for education purposes, reducing the cost, and by ensuring that the green energy demand is met, enhancing the security of the information, as well as easing the maintenance and operation of the system.

Several educational establishments in the United States have recognized the ability of cloud computing to improve efficiency, cut on costs and ensure convenience within the educational sector. For some institutions of higher learning, the high computing power associated with cloud computing is welcome for purposes of research. The cloud services have already been adopted within the education sector [5], with many learning institutions starting their movement towards cloud computing by having their student email system being provided by outsiders, with the most common one being Microsoft live@edu which is browser-based, and provides student access to email, office package, as well as skydrive. Live@edu is prevalent as it ensures that students can access Microsoft products from anywhere without necessarily having to purchase them [6]. Another one of the various education based cloud services is the Google App Education (GAE). Features from Google including Talk, Mail and Docs provide similar benefits as those of its counterpart Microsoft, hence enhancing the online student collaboration and their learning experience [7]. Learning institutions are further utilizing the lower level cloud computing services to store data. Further, cloud computing acknowledges the importance of M-learning by utilizing mobile devices as a way of enriching research through online collaboration which will ensure sharing of information and environment, hence leading to good quality work and success in research.

Cloud computing is also used in the education sector for hosting learning management systems (LMSs) e.g. Moodle and Blackboard within the cloud. Most institutions outsource the providers of the LMSs due to the high costs involved in the establishment and maintaining such systems [9]. In todays world e-learning has been adopted at different education levels including training for firms, lifelong learning, as well as in academic units; E-learning solutions range from commercial to open-source. There are two main entities of the e-learning system including trainers and students. The students get to access exams, courses, and can relay their assignments online, whereas the trainers can relay tests manage courses and evaluate homework and assignments for the students and the two parties can communicate with one another [8].

It is not possible for the e-learning solutions to ignore the current trends associated with cloud computing. Using cloud (SaaS) applications, it is possible for both teachers and students to access their individual data using a web browser from a computer or mobile phone at school, home, library or from any other place, hence ensuring efficient collaboration, communication and exchange of shared documents, notes, as well as contacts among other data. By using cloud applications, it means that teachers and students can be mobile and at the same time achieve their learning objectives by using portable laptops, and other devices that are interconnected. For example, it is possible for teachers and students to participate in class without the school premises. Below is an example [13] of how the utilization process of the various cloud services to the advantage of the management, students, teaching staff, software developers, as well as the research staff., The demand for IT services is a need that is directed towards the IT department (as illustrated in figure 1) whose duties include:

- To provide staff and students with the required and relevant software e.g. (operating systems, email accounts, malware detectors) and hardware e.g. (servers and PCs)
- To provide post-graduate students and researchers with the necessary hardware and software necessary to conduct experiments requiring high levels of computation and processing.
- To provide software and web developers with the necessary tools to host and write web applications.

Various aspects of the above mentioned elements can be easily transferred to cloud as showed in figure 2. For example lecturers, administrative staff as well as students may be compelled to use services from service providers of LaaS and SaaS clouds, which will ideally be accessed via clients. The software that may be launched by such groups remains to be on the SaaS servers and can be accessed online. In case additional hardware or disk space is required, it can be provided by the provider of IaaS online. Developers are also able to use all software necessary for their development from online sources, and all the necessary hardware required for hosting applications via the PaaS provider. Additionally, researchers with projects in need of high processing power can have it done via the IaaS provider.

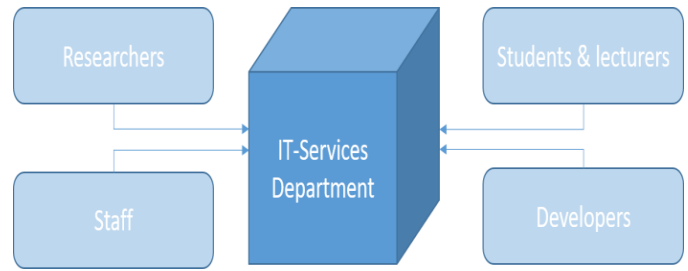

Fig. 1. The main structure of the IT services users in any institution.

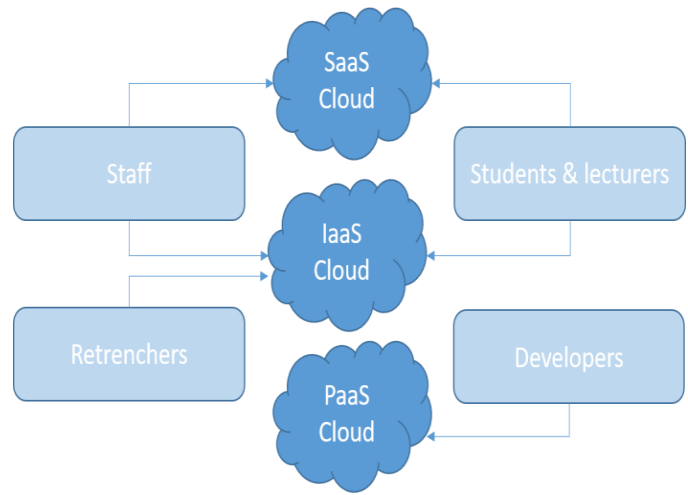

Fig. 2. The main structure of the IT services users in any institution using Cloud Computing.

Universities and colleges constantly seek to upgrade their IT hardware and software as a way of attracting students and keeping up with the rapidly changing technological environment. Cloud computing can enable such institutions to accomplish such objectives at an affordable cost. Moreover, by shifting the responsibility of maintaining and managing hardware and software to external providers ought to minimize costs because of the reduced IT labor costs [13].

Cloud computing can also be used in resource library where student can learn from. A library building is considered to be a long-term project, which calls for long-term planning to ensure sustainable and healthy development. The research and teaching resources need to be the main objective of building such a library, and its construction ought to follow the laid out planning process including division of labor with a basic focus on the principles that guide mechanism innovation and making proper use of the available resources. There is need to classify, organize, as well as to absorb the current excellent results using services of cloud computing coupled with innovation and scientific quality in the construction to ensure that the library is properly equipped with the necessary teaching and research resources. A teaching library that uses cloud computing platform is characterized by the following:

- Fast and stable. This means that it is powerful, fast and convenient in its search capabilities.
- Its specifications for data storage categories are clear: classification specifications and standardization are favorable to build a sustainable repository.
- Resources are safe in the sky: data and all information are stored in the cloud where they are safe. Thus, there is no need to have a local backup.
- The resource platform has an interface that is user friendly. All users irrespective of their level are able to access the resources easily and fast. This has ensured that educational resources do achieve optimum sharing ability as well as openness [15].

According to Carroll and Kop, Cloud computing has the ability to enhance new interface metaphors, as well as fresh ways of thinking in relation to learning and design experiences [10]. However, it is necessary to recognize cloud computing capabilities so as to find its application and use in education. According to Gartner, only 4% of the educational sector stakeholders utilize cloud computing in carrying out various educational services which means the cloud computing in educational sector still in the beginning [11]. Another study indicated that 12% of the interviewed respondents were conversant with cloud computing within the education sector, whereas 88% of the respondents agreed that there is need to implement cloud computing in the education sector [12].

By moving from traditional teaching models and adopting cloud computing would bring positive change to learning institutions though this technology may not be embraced by all since some stakeholders including teachers may feel that they may be replaced by e-learning systems though this ought not to be the case because teachers have an important role to play in education and they cannot be replaced.

## 3.1 Benefits of Cloud Computing in Education

The key benefit of Cloud computing in education is financial: Just like in other industries, learning institutions will be able to reduce costs when it comes to maintaining the IT infrastructure including licensing, energy consumption, technical labor, as well as hardware due to virtualization that comes with cloud computing. The user experience is highly enhance with the adoption of cloud computing. In this case the users will include teachers and students. Learning content can be retrieved from a central point and can be accessed from anywhere anytime instead of being relayed from local servers only, hence administration and management gets streamlined. Centralized learning content ensures that the staffs focus is on providing high quality learning experience rather than struggling with a system that is inefficient.

The learning needs of students in the current generation are quite different from those of the past generations. As a result of the changing times, students in the current generation prefer the use of technology and its various applications. Cloud computing provides students with fast access to core course materials and connects them with one another. Following are some of the benefits of adopting cloud computing by education institutes;

- It allows the users to use their personal workspace.
- It ensures that learning and teaching becomes more interactive
- It has high processing power for carrying out various operations
- There is no need of backups to store information since it is stored in the cloud.

Despite many benefits accruing to the education sector from cloud computing, there are potential risks involved and they include concerns about privacy, security, as well as matters relating to technological performance and reliability. Cloud computing also provides personal portfolio, digital environment for learning, as well as self-service options that are web-based to both staff and students. Other benefits of cloud computing for students include;

- Omnipresence of online applications.
- Flexibility: This allows for the creation of learning environments that are structured.
- It supports mobile learning.
- scalability.

Cloud computing provides a good alternative the education environment being used today. The various advantages that come with cloud computing may be the solution to various challenges in the education sector today. The following is a comprehensive list of advantages that come with the use of cloud computing within the education sector;

1) Institutions of higher learning can allow their technological infrastructure to be used by other firms as a way of enhancing research
2) The efficiency associated with cloud computing may aid learning institutions to keep pace with the growing need for energy cost and resources.
3) Cloud computing has an extended reach which enables universities to teach students in different and new ways, as well as ensure they can manage massive workloads and projects better.
4) It helps students appreciate new technology better when they join the global workforce.
5) Cloud computing exempts learning institutions from data management, and reduces costs and requirements that come with data security.
6) It provides services and online tools that ensure collaboration capabilities and secure communication.
7) It allows students and teachers to access, publish and share class calendars, documents, as well as web pages
8) Problems including insufficient infrastructure, lack of teachers, low rates of graduation, as well as tiny classrooms can be addressed by use of cloud computing.
9) Geographical distances will be bridged as people can study from anywhere
10) Institutions that lack adequate infrastructure can also provide education of high quality because they ought not to purchase software licenses, hardware, or incur implementation services costs.
11) Democratization of education is possible since cloud computing can be rapidly deployed by the institution.
12) Upgrades and maintenance ought to be easier. This is because the cloud enables rapid provision, acquiring, as well as deployment of new IT services, platforms, as well as applications.
13) IT capital expenditure is eliminated, hence a reduction in the overall cost outlay.
14) Service is available anytime any day as required by the user.
15) Good accessibility because service and data are available to the public.
16) It helps to move towards a greener world as it reduces the carbon footprint.
17) It is user friendly and can be used to manage large data quantity.

## 3.2 Challenges of Cloud Computing in Education

There are various challenges that come with cloud computing in education. Some of these challenges include security, data privacy, as well as insufficient network. Data handling, as well as privacy laws need not be taken flippantly. Some of the hindrances to the adoption of cloud computing in the education system include:

- **Security and Privacy:** This is a major concern among many institutions of higher learning [21, 22]. Cloud computing calls for the introduction of a third party who is the platform provider hence the privacy and security of data is hard to maintain.
- **Real Benefits:** Most institutions of higher learning are not yet convinced of the benefits that come with cloud computing. Such institutions are more concerned with their conventional IT portfolio and how to make cloud computing part of it. Hence, it is of essence for learning institutions to realize the benefits associated with cloud computing instead of just thinking of its potential. Thus there need to be set indicators for comparing performance and availability versus service level agreements (SLA), utilization, as well as costs.
- **Service Quality:** This is one of the reasons cited by learning institutions for not shifting to cloud computing. They argue that the SLAs stipulated by the providers of cloud services are insufficient when it comes to guarantying performance, availability, as well as scalability. Thus, without adequate guarantee, institutions of higher learning shy away from adopting cloud computing in running their operations.
- **Lack of adequate network responsiveness:** In case bandwidth of the network is not adequate, then it becomes impossible to deliver complex services through it. Most learning institutions lack adequate bandwidth, hence cannot adopt cloud computing affectively.
- **Integration:** Different applications require complex integration as to connect to the available on-premise applications, as well as cloud applications. This calls for the integration of existing university data structures and systems with cloud applications. Thus, there is need to have a quick, cost effective and simple way to connect university systems with cloud applications.

## 3.3 Guidelines To Successful Integrated Cloud Services

HP recently sponsored a worldwide IDC survey that monitored the experience of a group of cloud managers that was deemed to be proactive. According to the analysis done, five critical steps to ensure success in integrated cloud management were established [23];

1) Identify a plan which coordinates the firms modernization strategy, with its agenda for cloud infrastructure, as well as it SaaS agenda.

2) Assess the existing costs, and then come up with benchmarks for ongoing resource usage and application support.

3) Identify different opportunities that can be used to reduce cost and at the same time speed up the delivery of service through automation, as well as infrastructure provisioning.

4) Systems to integrate and monitor the performance of applications, as well as real-time planning of the capacity using automated provisioning solutions.

5) Security priorities and strategies need to be integrated in the development, launch and operations of the application.

Institutions that wish to acquire cloud services ought to have a selection criteria list, which they will give to cloud providers and have their response. The list should include aspects such as;

- **Functionality:** The functionality needed by users ought to be stated.
- **Technical issues:** The learning institution should conduct technical integration tests including automating and creating a user account on cloud to ascertain it functionality.
- **Platform:** The various platforms that support the applications need to be assessed.
- **User accessibility and experience:** Systems are not identical when it comes to providing user experience. If any other software needs to be installed other than the browser then it makes the system less attractive. Thus, it is important for the institution to ensure that its software resonates with the accessibility standards and guidelines provided before deployment of cloud services.
- **Contract:** The contract by the provider should be of a standard nature and ought to be closely studied.
- **Costs:** The actual cost to the higher learning institution may be of a considerable amount despite the cloud services cost being minimal.

## 3.4 Examples of Using Cloud Computing in Education

Educational institutions in the USA have recognized the impending role of cloud computing in improving the expenditure, competence and convenience for the education sector. Below are examples of cases that have incorporated cloud computing in education.

### 3.4.1 Eastern Michigan University

In 1999, the Eastern Michigan University (EMU) adopted the web-based online course content and class interface. Despite the undertaking being on the margins of cloud-computing, there was successive expansion and cloud supported course materials and support applications were available to learners. For the system to work efficiently in line with the vast needs of the study body and to improve on functionality, since the system relied entirely on cloud computing, there was need for successive updating of the original version. EMU-Online, a cloud-based educational application was implemented in 2010. It allows students to access threaded discussions, class email functions, gradebook access and shared documents through a web portal. All students relied on the application as their exclusive source of course information by 2014. In the same year, new developments facilitated students access to Google cloud suite directly in both the EMU-Online shell and the student records portal. The university is on a transition phase from EMU-Online to Canvas, which is also a cloud based application, with the same functionalities as EMU-Online but the interface is a bit user friendly [28].

### 3.4.2 University of California

The University of California (UC) viewed cloud computing to be appealing for use in one of its courses which was solely aimed at deploying and developing applications for SaaS. A donation from Amazon Web Services helped the UC course content move from being locally owned to the clouds, since many servers (needed for the course) could be easily obtained in the shortest time possible [18].

### 3.4.3 University of Westminster

The University of Westminster (UOW), with a student capacity of greater than 22,000 students in the UK has also embraced cloud computing. An outdated student email platform and a survey whose report showed that 96% of the student population was opting to set up personal email accounts which would provide by third party account where all the received emails could be automatically forwarded from their university account, sparked the universitys interest in cloud computing. In a quest for an alternative solution to the issues, they found GoogleApps (Educational Edition) which provided every student or staff member with a free email of 7.3GB of disk space, shared calendars and messaging. Another advantage of the package is that it allows users to retain their domain name, for instance, a user with justin@wmin.ac.ke as his email address is able to use this email address. In addition, the platform offers a set of productivity applications such as spreadsheet, word processing, with a collaborative functionality that allows users to shared documents remotely, a feature that is helpful to students in group assignments. After conducting pilot testing and consultation, the app, which has a 7.3GB email storage capacity, was launched for use during the 2008/9 academic year. A couple of problems that the students were facing was that, first, emails from the university were being treated as rogue, spam or bogus messages or even getting blocked once they were forwarded to their personal email accounts, thus students were not receiving urgent and key university emails. Secondly, network servers and email were experiencing storage based hurdles making students to resort to USB memory disks which were either lost or misplaced.

GoogleApps become the ultimate solution to spam emails and storage issues. It also gave the University an alternative of using friendly names instead of a students ID number for the email domain. Thirdly, students could access emails and save documents from their mobile devices. Lastly, the University was able to save up a lot of money since the costs of Google mail were nil. However, the university remains using the universitys old email system which is Exchange/Microsoft Outlook as the oficial staff email system. This was the university decision because they concerned about how transferring their data from their safe keep to a third party would have legal implications [13].

### 3.4.4 Kentuckys Pike County Schools

Kentuckys Pike County district introduced cloud computing to its schools to help in rationalizing running costs. ICC Technology Partner, an IBM subcontractor is responsible for managing the platform. Cloud computing has enabled the schools to revamp 1400 old computers, into full functional, that were idle waiting to be turned to scrap. The availability of the old computers on location to perform computer-based formative assessment was an advantage to the county since lack of enough computers was a potential problem. The availability of desktops was a bonus because cloud-computing was possible since with a cloud computing system a hard drive on a local computer is not needed. Processing is rather conducted at server level and not at the desktop level. The function of the desktop is acting as a dumb or conduit workstation that accepts software and processing capability sent from the cloud. According to the county, for half a decade, the expenditure for the hosted virtual desktop solution will be reduced to half the expenses of supporting the desktops on ground. Also, hosting the monitors in IBMs data center helped to evade further infrastructure and staffing expenses [14].

### 3.4.5 Florida Atlantic University

Florida Atlantic University, a public university, has a capacity of 29,000 students upwards and 170 degree programs. The university uses HyperV which is a hypervisor based server virtualization platform that merges workloads onto one server. By virtualizing its data center, university has been able to decrease IT expenses by U.S $600,000 and to deliver new IT services without additional staff. It was also possible for the university to run Blackboard on Linux in the HyperV setting, to deliver more performance and to simplify administrative work [20].

## 4 CLOUD COMPUTING IN E-LEARNING

Web based training (WBT). WBT involves learning the advancement of computer technologies to simplify tasks with the aid of preprogrammed software applications and virtual learning surroundings. E-learning is facilitated through a network enable computer that transfers the knowledge from internet sources to end users machines. Many institutions of higher learning are putting into place e-learning for their distance education students.

Despite the advantages of E-learning, many institutions do not have the adequate infrastructure and resources to implement state of the art e-learning. Blackboard and Moodle have come up with new versions of the base applications that are cloud oriented. This form of learning has also been adopted in continuous education, academic courses and company trainings due to its enhancement into the IT world through cloud computing, communication technology and mobile learning.

The benefits of e-learning to students include taking an online course, taking examinations, sending feedback, projects and homework. Trainers on the other, hand, can manage learning content, prepare and asses homework, tests and projects, offer feedback and have discussion forum with students.

The goal of modern distance education program is to achieve 4A, Anyone, Anytime, Anyplace and Anything. However, the current e-learning platform for the learner falls short of the 4A. Distance education offered on a cloud computing platform will improve the current situation. The availability of broadband internet access will make it easier for users to get all the necessary study materials, from anywhere by simply signing up and then logging into their accounts. A cloud computing enhanced web learning platform offers agile and flexible learning, a learning friendly environment and improved learning results.

Traditional e-learning networks are usually built, developed and maintained by the user institution. The expenses of obtaining equipment, developing and maintaining the system can be afforded by the user institutions, but they are rather costly. Transferring the building, maintenance, development and management to vendors, opening it up to numerous subscribers via the internet and allowing them use-on-demand and paying according to the total used servers would help to cut costs for the institutions and to achieve economies of scale to the vendor. With all the developments and advantages of cloud computing, it is hard for e-learning not to embrace cloud computing.

## 4.1 Advantages of using e-learning in the cloud

The benefits of an e-learning system in the clouds are numerous for the educational institutions dealing with them. These benefits include:

- Using the cloud applications on gadgets with Internet access helps to reduce costs since payment is only for applications and data and there is no need for equipment.
- There is enhanced performance in cloud computing because most applications have a large number of the processes and applications already available.
- Successive updates on the applications are quick to run and they are usually available on the spot.
- The document formats are compatible and there is improved compatibility on the platform based on the software chosen to open the files available from the server.
- The platform allows students to access course materials, projects and assignments instructions, and feedback from instructors and exams and to hold discussions from a flexible and convenient location.
- Another benefit is that the platform enable the instructors to take care of the student personally via online course access, test and offering personalized feedback.
- Running an e-learning business transfers building, developing and maintenance costs from the educational institution to the supplier which reduces costs and increases economies of scales, respectively.

## 5 IMPACT OF CLOUD COMPUTING ON THE SAUDI EDUCATIONAL SECTOR

Day after day, Cloud computing in Saudi Arabia is becoming more visible. The ability of cloud computing to improve productivity and efficiency as per the needs and demands of an organization has seen the platform grow in terms of visibility and use. Universities in Saudi Arabia have been permitted by the government to provide higher education to all its citizens. The Saudi government has injected a lot of effort and resources to enhance the educational system through development initiatives and putting up more universities. The Ministry of Higher Education (MOHE) has been asked to create, regulate and execute law regarding the higher education system in the country.

In Saudi Arabia, the enrollment numbers being estimated to be about a million yearly. The massive enrollment, need to reduce financial costs through coordination of approved programs, training methods and production of educational materials and the scarcity of available and qualified faculty members has seen the potential of e-learning grow. Efforts have been made by both governmental and non-governmental organizations towards implementing an up-to-date e learning experience for students. The Blackboard, which happens to be the most widely used education system is being used in the country to assist in blended learning.

The premeditated development of e-learning looks promising in terms of efficiency and delivery, reach and learn effectiveness. In a bid to provide state of-the-art educational facilities, the government has richly invested in the IT sector to obtain infrastructure to upgrade the systems.

Following, three fundamentally new impacts that must be factored into the educational system has been discussed:

- **Low Cost and Free Technology:**
  The IT sector has seen tremendous growth of low cost and free products that are suited for varied use such as collaborating, publishing, social interaction, content creation, editing or computing. With access to a web browser, most of the technology is easily accessible and even free, making them easily available for students.

- **Content Growth:**
  The traditional system of content being available from relatively known avenues such as encyclopedias and textbooks has become outdated. Today, the content is only a few clicks away with exponential growth in the available content, since anyone can contribute.

- **Collaboration:**
  The traditional face-to-face or via telephone communication have been replaced by technology that facilitates communication and collaboration with others. Dynamic teams and interactive collaboration area now easy to do with cloud computing.

Institutions of higher learning or business in Saudi Arabia can make most use of cloud computing for e-learning due to its reliability, efficiency, cost, security and portability.

## 6 DISCUSSION AND FUTURE WORK TO MOVE INTO CLOUD COMPUTING FOR SAUDI ARABIA INSTITUTIONS

There are generally two types of data under discussion with regard to any university. The first category encompasses all educational materials such as course specification, course material, curriculum, assignments and student projects that are done by that specific university. This form of data is considered as institutional data. Secondly, the other type of data is referred to as University internal which comprises all the data that each institution of higher learning own. The main aim of this type of data is for the application by the relevant university to conduct university official activities. In the event that the institutional data is shared with other institutions of higher learning with the authority and control of the relevant bodies, there would be the aspect of eliminating effort duplication. At the same time, the students and their professors would benefit a lot. This is contrary to university internal data as this data is private and cannot be shared with other institutions.

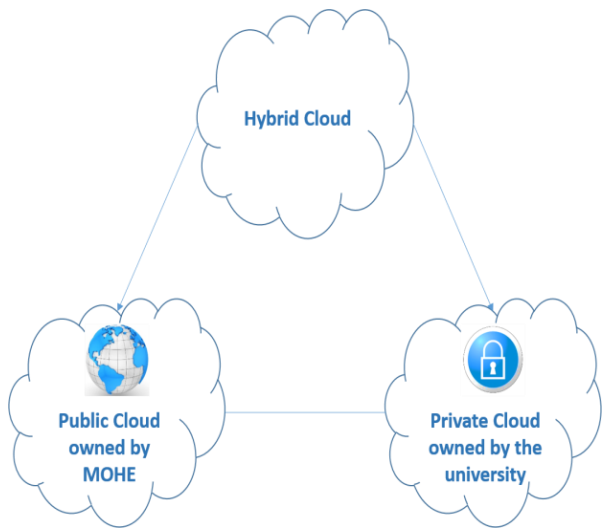

Fig. 3. The proposed hybrid cloud model.

In this case, the proposed model as illustrated in figure 3 is based on those two categories of data types. It is also suggested that a hybrid cloud model would simplify it for higher institutions of learning in Saudi Arabia for the purposes of sharing research activities and other aspects of knowledge. Taking into consideration that all institutions of higher learning in Saudi Arabia are governed by MOHE guidelines, there is the need for a public cloud owned by MOHE that contains institutional data generated by the institutions. It is therefore prudent for the other type of data to be hosted in a private cloud that is owned by the relevant university or college. The hybrid cloud will be mandated to host both the private and the public cloud. For the purposes of protecting the integrity of the data stored in the public cloud, MOHE is charged with the responsibility to ensure that issues of data plagiarism are averted. Universities should also prohibit the sharing of data on private cloud with other institutions but can grant access to the data to its users depending on their privileges.

The government can therefore apply this model to reduce a substantial costs in technology infrastructures in higher institutions of learning and provide other avenues of sharing knowledge and research among institutions of learning.

# 7 CONCLUSION

The economic growth of any country is usually maintained and enhanced by education in terms of quality and level. Cloud computing is an exciting development in todays education system. It offers students and administrative workers to an avenue to access different applications and resources through web pages easily, at minimal costs and quickly. Many organizations have experienced reduced running costs, improved efficiency and functionality due to the gradual removal of costs incurred from licenses, managing, hardware and software. The flexible aspect of cloud computing relieves IT staff of maintenance costs and duties, thus eliminating high operational costs and disaster recovery risks and its costs.

Cloud computing creates a universal platform with simplified scalability. Therefore it will be vital for schools and individuals to shift to the cloud, to experience the cheap and convenient avenue to information and technological services, especially the benefits and abilities, such as access to complex applications, minimal costs of cloud data storage, scalability and flexibility of an e-learning platform that is cloud computing enabled. However, e-learning integration in universities has some shortcomings that must be considered before the adoption and integration of the system.

The above discussion has shed a light on the adoption process of cloud computing into the education sector with deployment guidelines. The few challenges that are likely to be experienced can be resolved by new and better policies and techniques. Finally, a proposed hybrid cloud model for higher institutions of learning in Saudi Arabia has been discussed. As a future work, we need to validate this model and investigate different security aspects in order to implement it for higher learning institutions.